\documentclass[aps,prd,nofootinbib]{revtex4-1}

\usepackage{graphicx}
\usepackage{enumitem}
\usepackage{subfigure}
\usepackage{amssymb}
\usepackage{amsmath}
\usepackage{bbm}
\usepackage{color}

\begin{document}

\title{Relating the inhomogeneous power spectrum to the CMB hemispherical
anisotropy}

\author{Pranati K. Rath$^{1,}$\footnote{Email:pranati@iitk.ac.in}}
\author{Pavan K. Aluri$^{2,}$\footnote{Email:aluri@iucaa.ernet.in}}
\author{Pankaj Jain$^{1,}$\footnote{Email:pkjain@iitk.ac.in}}
\affiliation{$^1$ Dept. of Physics, Indian Institue of Technology Kanpur, Kanpur - 208016, India}
\affiliation{$^2$ IUCAA, Post Bag 4, Ganeshkhind, Pune University Campus, Pune - 411007, India}

\begin{abstract}
We relate the observed hemispherical anisotropy in the cosmic microwave 
radiation data to an inhomogeneous power spectrum model. The hemispherical
anisotropy can be parameterized in terms of the dipole modulation model. 
This model leads to correlations between spherical harmonic coefficients  
corresponding to multipoles, $l$ and $l+1$.
We extract the $l$ dependence of the dipole modulation amplitude, $A$,
by making a fit to the WMAP and PLANCK CMBR data. 
 We propose an inhomogeneous 
power spectrum model and show that it also leads to correlations between
multipoles, $l$ and $l+1$. The model parameters are determined by making
 a fit to the data. The spectral index of the inhomogeneous power spectrum
is found to be 
 consistent with zero.
\end{abstract}


\maketitle
\section{Introduction}
The cosmic microwave background radiation
(CMBR) shows a hemispherical power asymmetry 
with excess power in the southern ecliptic
hemisphere compared to northern ecliptic hemisphere \cite{Eriksen2004,
Eriksen2007,Erickcek2008,Hansen2009,Hoftuft2009,Paci2013,Planck2013a,
Schmidt2013,Akrami2014}. The signal
 is seen both in WMAP and PLANCK data and indicates a potential violation of the cosmological principle.
The hemispherical anisotropy can be parametrized phenomenologically by the 
dipole modulation model \cite{Gordon2005,Gordon2007,Prunet2005,Bennett2011} 
of the CMBR temperature field, which is given by,
\begin{equation}
 {\bigtriangleup T}(\hat n) = f(\hat n) \left(1+A \hat \lambda \cdot \hat n \right)\,,
\label{eq:dipole_mod}
\end{equation}
where $f(\hat n)$ is an intrinsically isotropic and Gaussian random field and,
$A$ is the amplitude of modulation along the direction $\hat \lambda$.
Taking the preferred direction along the z-axis, we have $\hat\lambda\cdot
\hat n=\cos\theta$. 
Using the WMAP five year data, the dipole amplitude for $l \le 64$
was found to be $A=0.072\pm0.022$ and the dipole direction, 
$(l,b) =(224^o,-22^o)\pm24^o$,  
in the galactic coordinate system 
 \cite{Hoftuft2009}. The 
PLANCK results \cite{Planck2013a} confirmed this anisotropy with a significance 
 of $3\sigma$ confidence level. 
A dipole amplitude,  $A=0.073\pm0.010$, in the direction
of $(l,b)=(217^o,-20^o)\pm15^o$ was found in PLANCK's SMICA map, which is also
seen (nearly with same amplitude and direction) in other PLANCK provided clean
CMB maps viz., NILC, SEVEM and COMMANDER-RULER maps. Hence
the results obtained by WMAP and PLANCK observations
are consistent with one another. 
There were indications that the hemispherical anisotropy might extend
to multipoles higher than $64$ \cite{Hoftuft2009,Hansen2009}, however, 
the effect is found to be absent beyond  $l\sim 500$
 \cite{Donoghue2005,Hanson2009}.  
The large scale structure surveys also do not show any evidence for
this anisotropy \cite{Hirata2009,Fernandez2013}. This suggests that any model
 which attempts to 
explain these observations 
should display a scale dependent power  \cite{Erickcek2009}
which should lead to a negligible effect
at high$-l$.

There also exist other 
observations which indicate a 
potential violation of the cosmological principle  \cite{Jain1999,Hutsemekers1998,Costa2004,
Ralston2004,Schwarz2004,Singal2011,Tiwari2013}. 
Many theoretical models,
which aim to explain the observed large scale anisotropy, have been proposed 
\cite{Berera2004,ACW2007,Boehmer2008,Jaffe2006,Koivisto2006,Land2006,
Bridges2007,Campanelli2007,Ghosh2007,Pontezen2007,Koivisto2008,Kahniashvili2008,
Carroll2010,Watanabe2010,Chang2013a,Anupam2013a,Anupam2013b,Cai2013,Liu2013,
Chang2013b,Chang2013c,Aluri13,Mcdonald2014,Ghosh2014,Panda14}. 
It has also been suggested that this anisotropy
may not really be in disagreement with the inflationary Big Bang cosmology,
which may have a phase of anisotropic and inhomogeneous 
expansion at very early time.
The anisotropic modes, generated during this
early phase may later re-enter the horizon
\cite{Aluri2012,Pranati2013a} and lead to the observed
signals of
anisotropy.

In a recent paper 
\cite{Pranati2013b}, 
we showed that the dipole modulation model, given in 
Eq. \ref{eq:dipole_mod}, leads to several implications for CMBR. 
The CMBR temperature field may be decomposed as,  
\begin{equation}
{\bigtriangleup T}(\hat n) = \sum_{lm}a_{lm}Y_{lm}(\hat n)\,.
\end{equation}
If we assume statistical isotropy,
the spherical harmonic coefficients must satisfy,
\begin{equation}
\langle{a_{lm}a^*_{l'm'}}\rangle_{iso} = C_{l}\delta_{ll'}\delta_{mm'}\,,
\label{eq:corr_iso}
\end{equation}
where, $C_l$ is the angular power spectrum. 
However in the presence of dipole modulation,
statistical isotropy is violated and one finds
 \cite{Pranati2013b}, 
\begin{equation}
\langle{a_{lm}a^*_{l'm'}}\rangle = \langle{a_{lm}a^*_{l'm'}}\rangle_{iso} +\langle{a_{lm}a^*_{l'm'}}\rangle_{dm}\,, 
\label{eq:corrdm}
\end{equation}
where, $\langle{a_{lm}a^*_{l'm'}}\rangle_{iso}$ is the correlation given 
in Eq. \ref{eq:corr_iso} and the anisotropic \emph{dipole modulation} term 
 can be expressed as,
\begin{equation}
 \langle{a_{lm}a^*_{l'm'}}\rangle_{dm} = A\left(C_{l'}+
                                            C_l\right)\xi^{0}_{lm;l'm'}\,. 
\label{eq:corr_aniso}
\end{equation}
Here, $\xi^{0}_{lm;l'm'}$ is given by,
\begin{eqnarray}
  \xi^{0}_{lm;l'm'} 
& \equiv &\int d\Omega Y_l^{m*}(\hat n)Y_{l'}^{m'}(\hat n)\cos{\theta} \nonumber\\
&= &\delta_{m',m}\Bigg[\sqrt{\frac{(l-m+1)(l+m+1)}{{(2l+1)}{(2l+3)}}}\delta_{l',l+1} \nonumber\\
&&
+\sqrt{\frac{(l-m)(l+m)}{{(2l+1)}{(2l-1)}}}\delta_{l',l-1}\Bigg]\,.
\label{eq:xillprime}
\end{eqnarray}
Hence the modes corresponding to the multipoles, $l$ and $l+1$ are correlated.
We thus define a correlation function 
\cite{Pranati2013b}, 
\begin{equation}
 C_{l,l+1} = \frac{l(l+1)}{2l+1}\sum_{m = -l}^{l} a_{lm}a^*_{l+1,m}\,.
\label{eq:corrl_l+1}
\end{equation} 
Here the factor $(2l+1)$ arises in the denominator in order to obtain an
average value of the correlation for a particular $l$. Furthermore
we multiply by $l(l+1)$ since for low $l$ the power $l(l+1) C_l$ is 
approximately independent of $l$. Using Eq. 
\ref{eq:corr_aniso} we deduce that
with this factor the correlation
$C_{l,l+1}$ would be roughly equal for different $l$ values.  
Analogously, the signal of hemispherical asymmetry is also 
observed in the variable $l(l+1) C_l$ \cite{Eriksen2004,Eriksen2007}. 
We define the statistic, $S_H(L)$,
by summing over a range of multipoles, 
\begin{equation}
S_H(L) = \sum_{l = l_{min}}^{L} C_{l,l+1} \,.
\label{eq:SH}
\end{equation}
We point out that if the factor $l(l+1)$ was not inserted in Eq. 
\ref{eq:corrl_l+1} then the statistic will be dominated by a few low $l$
multipoles. 
The final value of the data statistic is obtained by  
maximizing it over
the direction parameters. We make a search over the direction parameters 
in order to maximize the value of the statistic, $S_H(L)$. 
The resulting statistic is labelled as $S_{H}^{data}$.  
The corresponding direction gives us the preferred direction, $\hat \lambda$,
defined in Eq. \ref{eq:dipole_mod},
  which corresponds to the choice of
z-axis.

In this paper, our objective is twofolds:
\begin{enumerate}
\item We first update our results in Ref.~\cite{Pranati2013b} for the estimate
of the effective value of the dipole modulation parameter, $A$, as a function
of $l$. 
The multipole dependence of $A$ in Ref.~\cite{Pranati2013b} was
extracted using the dipole power of the temperature squared field.
In the currrent paper we instead use the statistic, $S_H(L)$,
which is a much more sensitive probe of
$A$ in comparison to
temperature square power. In particular, this statistic leads to a
higher significance of the signal of dipole modulation.
Furthermore in Ref.~\cite{Pranati2013b}, we
 used COM-MASK-gal-07 for the PLANCK data. In the
present paper we use the $KQ85$ mask for WMAP's nine year ILC 
map and the more reliable, 
CMB-union mask for SMICA, also known as the U73 mask.

\item We also propose a general inhomogeneous primordial power spectrum model.
This is presented in the next section, where we argue that the inhomogeneous
contribution to the power spectrum in any
model must reduce to ours, as long as it 
is small. The assumption of small inhomogeneity is reasonable and also
supported by our fit to data. An inhomogeneous power model is expected
to induce an anisotropy in the CMBR as well as in other cosmological 
observations. 
We show that our model leads
to correlations among the spherical harmonic coefficients similar to those
implied by the dipole modulation model. This allows us to 
relate the inhomogeneous primordial 
power spectrum with the 
dipole modulated temperature field and hence the hemispherical anisotropy.
 We determine the primordial power spectrum
which leads to the hemispherical power asymmetry observed in the CMB 
temperature data. 
\end{enumerate}

The observed 
anisotropy is parametrized by the two point correlations given by 
Eq. \ref{eq:corr_aniso},
or equivalently the statistic, $S_H(L)$. We first determine the data statistic,
$S_{H}^{data}$, as defined above after Eq. \ref{eq:SH}.  
We next determine the amplitude, $A$, of the
 dipole modulation model given by Eq. \ref{eq:dipole_mod}, by making a fit to data. 
We perform this analysis for the entire multipole range $l=2-64$ and
also, independently, 
 over the three multipole ranges, $l=2-22, 23-43, 44-64$. This allows
us to determine
the variation of $S_H^{data}$ and hence the dipole modulation amplitude, $A$,
 with the multipole bin. 
We determine the value of $A$ by simulations, as explained in section 
\ref{sec:dataanalysis}. The upper limit on $l$ ($l\le 64$) is imposed since
the amplitude of the dipole modulation using hemispherical analysis,
 has been studied most thoroughly only in this range 
 \cite{Hoftuft2009,Planck2013a}.  
As discussed in the next section, the theoretical analysis is also simplest
for low-$l$ modes. In this case we can make a simple approximation for
the transfer function. At higher $l$ the transfer function is not as simple 
 and the computation of the correlation, $\langle a_{lm} a^*_{l'm'}
\rangle$, in the presence of inhomogeneous and/or anisotropic is 
significantly more complicated.  
Furthermore, as we shall
see, the contribution due to detector noise is negligible in the range,
$l\le 64$. 
This leads to considerable simplification in our analysis. Simulations at 
 higher $l$ also require much higher computation time. Finally, as we
have discussed above, the signal of dipole modulation is expected to die
down for values of $l$ beyond a few hundred
 \cite{Donoghue2005,Hanson2009,Hirata2009,Fernandez2013}.  
We hope to extend our analysis
to higher values of $l$ in a future publication.   

We finally determine the 
inhomogeneous power spectrum by making a fit to the data 
statistic, $S_{H}^{data}$.  
 This calculation
is also performed over the entire multipole range, $l=2-64$,
and then repeated over the three multipole bins mentioned above.
 Hence we determine the magnitude as well as the wave number, $k$,
 dependence of 
the power spectrum.   
The possibility of inhomogeneous power has been considered earlier,
for example, in Ref.~\cite{Erickcek2008,Hirata2009,Carroll2010,Gao2011}.
However our analysis and results differ from these earlier papers.   

\section{Theory}
The temperature fluctuations, ${\bigtriangleup T(\hat n)}$, arise due to 
 primordial density perturbations, $\delta(\vec k)$. They can be expressed as, 
\begin{equation}
\frac{\bigtriangleup T}{T_0}(\hat n) = \int d^{3}k \sum_{l}(-i)^{l}(2l+1)\delta(k)\Theta_l(k) P_l(\hat k \cdot \hat n)\,,
\label{eq:DToverT}
\end{equation} 
where $P_l(\hat k \cdot \hat n )$ are the 
Legendre polynomials and $\Theta_l(k)$ the transfer function. 
On large distance scales, i.e. low$-l$,   
the main contribution to the temperature fluctuation arises
 from the Sachs-Wolfe effect. Hence we  
approximate the transfer function as $\Theta_l(k) =\frac{3}{10}j_l(k\eta_0)$
\cite{Gorbunov2011}, where $j_l$ is the spherical Bessel function and $\eta_0$
 the current conformal time. 
A more detailed analysis is postponed to future work.
Let $\tilde\delta(\vec x)$ represent the density fluctuations in real space.
The two point correlation function, $F(\vec\Delta,\vec X)$, in real space
can be expressed as,
\begin{equation}
F(\vec\Delta,\vec X) = \langle\tilde\delta(\vec x) \tilde\delta(\vec x')\rangle\,, 
\end{equation} 
where we have defined the variables, $\vec \Delta= \vec x-\vec x'$ and
$\vec X= (\vec x+\vec x')/2$. The corresponding correlation function  
of the Fourier transform, 
$\delta(\vec k)$, may be expressed as, 
\begin{equation}
\langle \delta(\vec k)\delta^*(\vec k')\rangle = \int {d^3X\over (2\pi)^3}
{d^3\Delta\over (2\pi)^3} 
e^{i(\vec k+\vec k')\cdot \vec \Delta/2}
e^{i(\vec k-\vec k')\cdot \vec X} F(\vec\Delta,\vec X)\,.
\label{eq:corr1}
\end{equation} 
If we assume that $F(\vec\Delta,\vec X)$ depends only on the magnitude
$\Delta\equiv |\vec \Delta|$, we obtain the standard form of the power spectrum,
$\langle \delta(\vec k)\delta^*(\vec k')\rangle=P(k)\delta^3(\vec k-\vec k')$,
where, $k\equiv |\vec k|$.  
An inhomogeneous model of power spectrum must necessarily depend on
$\vec X$ in real space. Hence for such a model $F(\vec\Delta,\vec X)$
cannot be independent of $\vec X$. 
Our procedure of obtaining an inhomogeneous power spectrum in Fourier space by 
using guidance from the real space power spectrum is similar to that 
followed in Ref. \cite{Carroll2010}. However the precise model we use is 
different from theirs.

We next assume a simple inhomogeneous model of $F(\vec\Delta,\vec X)$.
It is simplest to consider an inhomogeneous model in order to obtain
the observed signal of hemispherical anisotropy or dipole modulation. 
An anisotropic model leads to correlations between multipoles, $l$ and
$l+2$ \cite{Abramo2010}, 
in contrast to the prediction based on the dipole modulation model.
Hence it does not lead to the observed anisotropy.  
An inhomogeneous model of power spectrum in Fourier space must show 
dependence on the position vector, $\vec X$, in real space.  
We assume that our inhomogeneous model
 has a very mild dependence on $\vec X$. Hence we can make a
Taylor expansion about any chosen origin and keep only the leading order
term. We can, therefore, express $F(\vec\Delta,\vec X)$ as,
\begin{equation}
F(\vec \Delta,\vec X) = F(\vec \Delta,0) + X_i {\partial\over \partial X_i}F(\vec\Delta,\vec X)\Bigg|_{\vec X=0} 
\label{eq:FDelta_1}
\end{equation}
The Taylor expansion is justified if the second term is significantly
smaller than the first. As we shall see, this expectation is validated
in our numerical results.  
The first term on the right hand side corresponds to the standard 
homogeneous and isotropic power spectrum. Hence it should be a function only 
of the magnitude, $\Delta$. We parametrize it as, 
\begin{equation}
F(\vec\Delta,0) \equiv f_1(\Delta)   
\end{equation}
The second term on the right hand side in Eq. \ref{eq:FDelta_1} represents
the contribution to power spectrum due to inhomogeneity. 
We parametrize the derivative term as,
\begin{equation}
{\partial\over \partial X_i}F(\vec\Delta,\vec X)\Bigg|_{\vec X=0} 
\equiv  \hat \lambda_i   \left({1\over \eta_0}f_2(\Delta)\right)
\end{equation}
where $f_2(\Delta)$ depends only on the magnitude, 
$\Delta$, 
 $\hat \lambda$ is a unit vector and we have introduced a factor $\eta_0$
in order to make the function $f_2(\Delta)$ have same dimensions as 
$f_1(\Delta)$. 
We point out that 
since $\vec X$ has dimensions of length,  
 $\vec X/\eta_0$ is dimensionless. Here $f_2(\Delta)$ depends only on
the magnitude $\Delta$, since otherwise the model will give an additional
anisotropic contribution to the power spectrum beside the inhomogeneity. 
We finally obtain,
\begin{equation}
F(\vec \Delta,\vec X) = f_1(\Delta) + \hat\lambda\cdot\vec X \left({1\over \eta_0}f_2(\Delta)\right)\,,
\label{eq:FDelta}
\end{equation}
It is clear that the inhomogeneous contribution to the power spectrum
in any model 
can be expressed in this form by making a Taylor expansion, provided the  
 inhomogeneity is small. This of course assumes that the model does
not contain any other source of
 violation of the cosmological principle besides the
presence of inhomogeneity.  
The smallness of the second term implies that $f_2(\Delta)$ is sufficiently
small so that it gives a contribution smaller than the first term. 

In applications to CMBR,
we need to compute the correlation
in Fourier space, given in Eq. \ref{eq:corr1}. 
We express the correlation as,
\begin{equation}
\langle \delta(\vec k)\delta^*(\vec k')\rangle = \tilde F_0(\vec k,\vec k')
+ \Delta\tilde F(\vec k,\vec k')\,, 
\label{eq:corr_def}
\end{equation}
where $\tilde F_0(\vec k,\vec k')$ 
and $\Delta\tilde F(\vec k,\vec k')$ 
are, respectively, the contributions due to the 
homogeneous and inhomogeneous terms in 
Eq. \ref{eq:FDelta}. The first term is given by,
\begin{equation}
\tilde F_0(\vec k,\vec k')  = P(k)\delta^3(\vec k-\vec k')\,, 
\label{eq:F0homo}
\end{equation}
where $P(k)$ is the Fourier transform of $f_1(\Delta)$. 
We choose $P(k)$, or equivalently $f_1(\Delta)$, to be the standard 
power spectrum in an homogeneous and isotropic model. 

We next need the Fourier transform of the inhomogeneous term. 
This calculation is facilitated by noting that,
\begin{equation}
\int {d^3X\over (2\pi)^3} X_i e^{i\vec k_-\cdot \vec X} = \int {d^3X\over (2\pi)^3}{\partial\over i\partial k_{-i}}
e^{i\vec k_-\cdot \vec X} = -i{\partial\over \partial k_{-i}}\delta^3(\vec k_-)\,,
\end{equation} 
where $\vec k_- = \vec k-\vec k'$. 
Hence we obtain, 
\begin{equation}
\Delta\tilde F(\vec k,\vec k') = -i\hat \lambda_i \left[ {\partial\over \partial k_{-i}}\delta^3(\vec k_-)\right]\  \ g(\vec k_+)\,, 
\label{eq:DeltatildeF}
\end{equation} 
where, 
\begin{equation}
g( k_+) = 
{1\over \eta_0}  \int {d^3\Delta\over (2\pi)^3}e^{i\vec k_+\cdot\vec \Delta} f_2(\Delta) 
\end{equation} 
and $\vec k_+ = (\vec k+\vec k')/2$.  
The inhomogeneous nature of $\Delta\tilde F(\vec k,\vec k')$ can be seen by
the fact that it is proportional to the derivative of the delta function. 
We see directly from Eq. \ref{eq:corr1}
that a homogeneous model of power spectrum
must be proportional to $\delta^3(\vec k-\vec k')$ since it must 
be independent of $\vec X$ in real space \cite{Carroll2010}. 

The spherical harmonic coefficients, $a_{lm}$, of the temperature
field can be expressed as,
\begin{equation}
a_{lm} = \int d\Omega Y^*_{lm}(\hat n) \Delta T(\hat n)\,.
\end{equation}
Using Eq. \ref{eq:DToverT}
and the identity, 
\begin{equation}
P_l(\hat n \cdot \hat n') = \frac{4\pi}{2l+1}\sum_{m} Y_{lm}(\hat n)Y^*_{lm}(\hat n')\,,
\end{equation}
we obtain,
\begin{equation}
a_{lm} = (-i)^l(4\pi T_0) \int d^3k \delta(k) \Theta_l(k) Y^*_{lm}(\hat k)\,. 
\end{equation}
We take the preferred direction, $\hat \lambda$ along z-axis, i.e. 
$\hat\lambda=\hat z$.
The two point correlation function of the 
spherical harmonic coefficients, $a_{lm}$, can be expressed as, 
\begin{equation}
\langle a_{lm}a^*_{l'm'}\rangle = (4\pi T_0)^2\int d^3k d^3k' (-i)^{l-l'}
\Theta_l(k)\Theta_{l'}(k') Y^*_{lm}(\hat k) Y_{l'm'}(\hat k')
\langle\delta(\vec k)\delta^*(\vec k')\rangle  \,.
\end{equation}
We parameterize this correlation as,
\begin{equation}
\langle a_{lm}a^*_{l'm'}\rangle = 
C_l\delta_{ll'}\delta_{mm'} + A(l,l')\,,
\label{eq:corraap}
\end{equation}
where $C_l$ is the standard power spectrum arising due to the homogeneous
term, given in Eq. \ref{eq:F0homo}, and  
$A(l,l')$ represents the anisotropic part arising due to the inhomogeneous 
term. Setting, $\Theta_l(k)=3j_l(k\eta_0)/10$, 
the isotropic part is given by,
\begin{equation}
C_l = (4\pi)^2 \frac{{9T_0}^2}{100}\int_0^{\infty} k^2dk j_l^2(k\eta_0)P(k)\,. 
\end{equation}

For the anisotropic part, 
using Eq. \ref{eq:corr_def} and Eq. \ref{eq:DeltatildeF}, we obtain,
\begin{equation}
A(l,l') =(4\pi T_0)^2 \int d^3 k d^3 k' (-i)^{l-l'+1}
\Theta_l(k)\Theta_{l'}(k') Y^*_{lm}(\hat k) Y_{l'm'}(\hat k')
g(k_+) \left({\partial\over \partial k_{-z}}\delta^3(\vec k_-)\right) \,.
\label{eq:Allp1}
\end{equation}
This can be computed by integration by parts. The details are given in 
Appendix A. The surface term vanishes and
we obtain,
\begin{equation}
A(l,l') = -(4\pi T_0)^2\int d^3k d^3k' (-i)^{l-l'+1}
\delta^3(\vec k_-)
g(k_+) {\partial\over \partial k_{-z}}
\left[\Theta_l(k)\Theta_{l'}(k') Y^*_{lm}(\hat k) Y_{l'm'}(\hat k')\right] \,.
\label{eq:Allp2}
\end{equation}
This leads to,
\begin{eqnarray}
A(l,l') &=& -(4\pi T_0)^2\int d^3k d^3k' (-i)^{l-l'+1} \delta^3(\vec k_-) g(k_+)\nonumber\\
&\times &\Bigg\lbrace\Theta_l(k)\Theta_{l'}(k') \left[  
{-1\over k\sin\theta_k}
Y_{l'm'}(\hat k')
{\partial\over \partial \theta_{k}} Y^*_{lm}(\hat k)+ 
{1\over k'\sin\theta_{k'}}
Y^*_{lm}(\hat k)
{\partial\over \partial \theta_{k'}} Y_{l'm'}(\hat k')
\right]\nonumber \\
&+ & Y_{l'm'}(\hat k')Y^*_{lm}(\hat k)\left[{\partial\Theta_l(k)\over \partial k}\Theta_{l'}(k'){k_z\over k} - 
{\partial\Theta_{l'}(k')\over \partial k'}\Theta_{l}(k){k'_z\over k'} 
\right]
\Bigg\rbrace\,,
\end{eqnarray}
where $\theta_k$ and $\theta_{k'}$ are the polar angles of the vectors
$\vec k$ and $\vec k'$ respectively.
After integrating over $k'$, we can express this as,
\begin{equation}
A(l,l') = A_1(l,l') + A_2(l,l')\,,
\end{equation}
where,
\begin{eqnarray}
A_1(l,l') &=&  (-i)^{l-l'+1} (4\pi T_0)^2\int d^3k\  g( k)
{\Theta_l(k)\Theta_{l'}(k)\over k\sin\theta_k} \nonumber\\
&& \times \left[ 
Y_{l'm'}(\hat k)
{\partial\over \partial \theta_{k}} Y^*_{lm}(\hat k)- 
Y^*_{lm}(\hat k)
{\partial\over \partial \theta_{k}} Y_{l'm'}(\hat k)
\right] \,,
\label{eq:A1llp}
\end{eqnarray}
and
\begin{eqnarray}
A_2(l,l') &=&  -(-i)^{l-l'+1} (4\pi T_0)^2\int d^3k\  g( k)
Y_{l'm'}(\hat k) Y^*_{lm}(\hat k){k_z\over k} \nonumber\\
&& \times \left[
\Theta_{l'}(k)  {\partial\over \partial {k}} \Theta_{l}(k)-
\Theta_{l}(k)  {\partial\over \partial {k}} \Theta_{l'}(k) 
\right]\,.
\label{eq:A2llp}
\end{eqnarray}
We compute $A_1$ by directly using the formula for spherical harmonics,
\begin{equation}
Y_{lm}(\theta,\phi) = N_{lm} e^{im\phi} P_l^m(cos\theta)\,,
\end{equation}
where,
\begin{equation}
N_{lm} = \sqrt{{(2l+1)\ (l-m)!\over 4\pi\ (l+m)! }}\,. 
\end{equation}
Integration over $\phi$ in Eq. \ref{eq:A1llp} leads to $\delta_{mm'}$. 
The conventions for the spherical harmonics, followed in 
HEALPIX \cite{Gorski2005}, differ
slightly from those used above. However this difference does not produce
any change in our theoretical predictions.
We, therefore, obtain,
\begin{equation}
A_1(l,l') =  (-i)^{l-l'+1}2\pi\delta_{mm'} (4\pi T_0)^2 N_{lm}N_{l'm} I_l
\int dkk\   g( k)
\Theta_l(k)\Theta_{l'}(k) \,,
\label{eq:A1llp1}
\end{equation}
where,
\begin{equation}
I_l = \int_{-1}^1 dx \left[P_l^m(x){d\over dx}P_{l'}^m -
P_{l'}^m(x){d\over dx}P_{l}^m \right] = -2 \int_{-1}^1 dx P_{l'}^m(x)
{d\over dx}P_{l}^m \,.
\label{eq:Il}
\end{equation}
This integral, for $l'=l+2n+1$, $n=0,1,2,...$, evaluates to \cite{Samaddar1974},
\begin{equation}
I_l = -2\left[\delta_{0,m}-{(l+m)!\over (l-m)!}\right]\,.
\label{eq:Ilvalue}
\end{equation}
for $m\ge0$. The corresponding result for $m<0$ can be easily determined
by using this identity. 
The remaining $k$ integral in $A_1(l,l')$ has to be performed numerically.
In the present paper we shall set, $n=0$, and study only the correlations
between $l$ and $l+1$ multipoles. The model, however, also predicts
higher order correlations, which require a more detailed study.

The integral $A_2(l,l')$ can be expressed as, 
\begin{eqnarray}
A_2(l,l') &=& -(-i)^{l-l'+1}(4\pi T_0)^2 \int dkk^2d\Omega  \cos\theta g( k) \nonumber \\
&&  \times \left[\Theta_{l'}(k) {d\over dk} \Theta_l(k) - \Theta_l(k) {d\over dk}
\Theta_{l'}(k)  \right] Y^*_{lm}(\hat k)Y_{l'm'}(\hat k) \,.
\end{eqnarray}
Using the identity,
\begin{eqnarray}
\int d\Omega \cos\theta Y^*_{lm}(\hat k)Y_{l'm'}(\hat k)
 &=& \delta_{mm'}\left[\sqrt{{(l-m+1)(l+m+1)\over (2l+1)(2l+3)}} \delta_{l',l+1} \right.  \nonumber \\
 && \left. + \, \sqrt{{(l-m)(l+m)\over (2l+1)(2l-1)}} \delta_{l',l-1}  \right] \,,
\label{eq:id1}
\end{eqnarray}
we obtain, for $l'=l+1$, 
\begin{eqnarray}
A_2(l,l') &=& -\delta_{mm'}
\sqrt{{(l-m+1)(l+m+1)\over (2l+1)(2l+3)}} 
(4\pi T_0)^2\nonumber\\
&\times&\int dkk^2 g( k)
\left[\Theta_{l'}(k) {d\over dk} \Theta_l(k) - \Theta_l(k) {d\over dk}
\Theta_{l'}(k)  \right]  \,.
\label{eq:A2final}
\end{eqnarray}
Hence we
find that the inhomogeneous power spectrum leads to a 
correlation between $l$ and $l\pm 1$. This may be compared to the correlations,
Eq. \ref{eq:corr_aniso}, 
we obtained from the dipole modulation model. 
The statistic $S_H$ corresponding to the theoretical correlation,
Eq. \ref{eq:corraap}, is denoted as $S_{H}^{theory}$.

We set the homogeneous power, $P(k) = k^{n-4}A_{\phi}/(4\pi)$, 
where the parameters,  
 $n =1$ and $A_{\phi} = 1.16\times 10^{-9}$ \cite{Gorbunov2011}. 
For the inhomogeneous term, we assume the form,
\begin{equation}
g(k) = g_0 P(k) {(k\eta_0)^{-\alpha}\over \eta_0}\,.
\label{eq:powergk}
\end{equation}
 Here $g_0$ and $\alpha$
are parameters. 
We have assumed that $g(k)$ also has a power dependence on $k$, which is
same as that of $P(k)$ up to the extra term $k^{-\alpha}$. The precise
combination $(k\eta_0)$ is chosen since it is dimensionless. As explained
above, the factor $\eta_0$ is inserted in order make $f_1(\Delta)$ and
$f_2(\Delta)$ have same dimensions. This implies that the constant $g_0$ is  
dimensionless. This constant parameterizes the amplitude of inhomogeneity
in the model.
Using $g(k)$, we can compute $A(l,l')$. 
We obtain the theoretical prediction for correlations between multipoles
$l$ and $l+1$ by performing the integrals on the right hand sides of
Eqs. \ref{eq:A1llp1} and \ref{eq:A2final}
numerically. These can be used to obtain the theoretical estimate of the
statistic for any value of the parameters, $g_0$ and $\alpha$.

\section{Data Analysis} 
\label{sec:dataanalysis}
We use the cleaned CMB maps, WMAP's nine year ILC map (here after WILC9) \cite{WMAP9yrILC}
as well as the SMICA, provided by the PLANCK team \cite{Planck2013b}. 
 For WILC9 map we use the $KQ85$ mask and in the case of SMICA map, we use the 
CMB-union mask to  eliminate the foreground contaminated regions. 
The masked portions are filled by simulated random isotropic CMBR data. 
The cleaned SMICA map and the corresponding mask are provided at a high
resolution with $N_{side}=2048$. Hence in this case we generate 
a \emph{foreground residual free} full sky CMB data map at this 
high resolution. Subsequently we downgrade this map  
to a lower resolution with $N_{side} = 32$ after applying appropriate 
Gaussian beam to smooth the mask boundary \cite{Pranati2013b}.
This procedure eliminates any breaks that might be introduced at the
boundary due to the random filling procedure.
The WILC9 map is available only at lower resolution
with $N_{side}=512$. Hence the corresponding full sky map is generated
only at this resolution and subsequently downgraded to a low resolution
map with $N_{side} = 32$. 
In this case, the simulated map also includes the contribution
due to detector noise. 
For PLANCK data, we do not include this contribution  
since the corresponding noise files are too bulky. 
As we shall see, for WMAP data,
 the detector noise gives very little contribution
over the multipole range under consideration. 
This provides a good justification for neglecting the detector noise in our
analysis. 
We also use the SMICA in-painted map, provided by the PLANCK team.
 This is a full sky map in which the masked regions have been 
reconstructed by the in-painting procedure \cite{Abrial2008,Inoue2008}.  

We use the three maps, described above, 
to determine the statistic, $S_H(L)$, over the multipole
range $2\le l\le 64$. This statistic is maximized by making a search over the
preferred direction parameters, i.e. the choice of our z-axis.  
Due to the random filling of the masked regions, we obtain different
results for different realizations for the full sky `data' map. Hence the 
resulting quantities, i.e. maximum value of $S_H(L)$ and the
corresponding direction, $(l,b)$,
are obtained by taking their average over 
100 such filled data maps for the case of WILC9 and SMICA maps.

Our filling of masked regions with random isotropic data is likely to 
generate some bias in our results. We estimate this bias through simulations. 
The final results are obtained after making a correction for this bias.
As we shall see this bias correction is relatively small.
This procedure of bias correction through simulations is analogous to,
for example, the procedure 
used in correcting for the residual foreground contamination in
 the ILC map by the WMAP science team \cite{Hinshaw2007}.   
We first simulate a full sky CMBR map which has same characteristics as
the real data. This is obtained by the following steps: 
\begin{enumerate}
\item We first generate a full sky random isotropic CMBR map. 
\item The full sky map is multiplied by the dipole modulation term, 
$(1+A\hat\lambda\cdot\hat n)$,
where the direction corresponds to that obtained by the WMAP
 five year analysis
\cite{Hoftuft2009} and the amplitude is a free parameter. 
This creates a full sky realization of a simulated
CMBR map
which displays dipole modulation, as seen in the real data.  
\item We next apply the same mask to this map as applied in the case 
of real data.  
\item Finally the masked portions are filled with randomly generated isotropic CMBR data.
\end{enumerate}
The above procedure generates a full sky realization of a map with 
same properties
as the real map, including the dipole modulation. 
We now choose the amplitude $A$ 
 to be such that the final map leads to a statistic, $S_H$, same as 
that seen in real data, $S_H^{data}$. 
We perform this simulation by generating 1000  
 samples for a particular $A$. The output, $S_H$, is determined by taking
the average over these samples. The error in $S_H$ is given by the variance 
over these samples. This process is repeated over a large number of
values of $A$ in order to determine that value which leads to $S_H$ closest
to real data. The final $A$, determined by this procedure, gives us
an estimate of the dipole modulation amplitude in real data, along with
its error.  

The final extracted value of $A$ is used to determine the full sky 
estimate of the statistic. We use this value in order to generate
1000 full sky realizations of simulated CMBR maps, which have dipole modulation
amplitude equal to the final extracted value of $A$. The statistic, $S_H$, 
computed by taking an average over these maps gives
us an estimate of the full sky, bias corrected, statistic. 
As we shall see, the bias correction is relatively small.

The significance of anisotropy is determined by comparing the 
maximum value of bias corrected $S_H(L)$ with that obtained from 4000 isotropic 
randomly generated full sky CMBR maps. 
The simulations used to estimate the significance from WILC9 map
include detector noise,
whereas those used for PLANCK data analysis ignore this contribution.
The significance is quoted in terms of the P-value, which is 
defined as the probability
that a random isotropic CMB map may yield a statistic larger than that seen 
in data. For each randomly generated map, we make a search over the
direction parameters in order to maximize $S_H(L)$. The maximum values of
4000 random samples are compared with the data statistic. The P-value
is estimated by counting 
 the number of times the statistic of random data exceeds the
observed data value. 
We point out that the significance is computed only as a consistency 
check.
The main purpose of the present paper is to extract the
dipole modulation amplitude and the parameters corresponding to the
inhomogeneous power spectrum. In any case, 
 as we shall see,
our estimate of significance 
is in reasonable agreement with the significance obtained
in literature \cite{Eriksen2004,Eriksen2007,Hansen2009} by hemispherical
analysis.

 The calculation is first performed 
over the entire range of multipoles, $2\le l\le 64$, and then 
repeated over the three multipole bins, $l=2-22, 23-43, 44-64$. 
Alternatively, we may determine $A$ by using the formula,
Eq. \ref{eq:corr_aniso}, for the correlation, 
 $\langle{a_{lm}a^*_{l'm'}}\rangle_{dm}$, which is applicable for the
dipole modulation model. We have verified that the value of $A$ obtained
by this formula agrees, within errors, with our 
direct simulations estimate.  

We finally determine the inhomogeneous power spectrum 
i.e. the parameters, $g_0$ and $\alpha$,
defined in Eq. \ref{eq:powergk}. We determine
the theoretical prediction for the statistic, $S_H(L)$, using 
Eq. \ref{eq:corraap} for the correlation between different multipoles. 
We first extract the value of $g_0$ for which the theoretical value of
$S_H$ matches the data value over the entire range of multipoles, $2-64$. 
In this case we set $\alpha=0$. We next extract the values of $g_0$
and $\alpha$ by making a fit to the data statistic in the three multipole
bins, $l=2-22,23-43,44-64$, using
the $\chi^2$ minimization procedure. Here $\chi^2$ is defined as, 
\begin{equation}
\chi^2 = \sum_L {\left[ S_{H}^{theory}(L)- S_{H}^{data}(L)\right]^2\over 
[\delta S_H^{data}(L)]^2} 
\end{equation}
where the sum is over the three multipole bins.  
Here $S_{H}^{theory}(L)$ is the theoretical estimate of the statistic 
 in a particular multipole bin  using the 
inhomogeneous model, 
$S_{H}^{data}(L)$ the bias corrected estimate  
and $\delta S_H^{data}(L)$
the corresponding error.  
In this case also we first set $\alpha=0$ and determine the best fit
value of $g_0$. Next we determine both of these parameters. 

\section{Results}

The extracted value of the data statistic, $S_{H}^{data}$, is given in 
Table \ref{tb:dipole1} for the maps WILC9, SMICA and SMICA in-painted. 
This is obtained by searching over all possible directions
in order to maximize $S_H(L)$. 
The corresponding direction
parameters are found to be in good agreement with those found by
using hemispherical anisotropy. For the case of WILC9 and 
SMICA we give both the raw and bias corrected values of 
$S_{H}^{data}$. 
 The anisotropy is found to be significant
roughly at $3\sigma$ confidence level, which is in reasonable agreement with the estimate
obtained by hemispherical analysis \cite{Eriksen2004,
Eriksen2007,Hoftuft2009,Planck2013a}. 
 In Table \ref{tb:dipole1}, the significance is given
in terms of the P-value, defined in section 3.   

For the case of WILC9 the random realizations used to fill the masked 
regions as well as to generate simulated data 
include contribution due to the detector noise. If this contribution
is not included, we find that the maximum value of $S_H(L)$,
without applying bias correction, is also equal to
$0.023\pm 0.007$. It is same as the value given in Table 
\ref{tb:dipole1} for WILC9.  
Hence the difference between the two cases is 
negligible. Furthermore the results of WMAP are in good agreement with
those obtained by using PLANCK data. Hence we do not expect detector
noise to be important in the multipole range of interest in this paper. 

We next determine the value of the dipole modulation parameter, $A$,
by comparing $S_{H}^{data}$ with the statistic 
obtained by the simulated dipole modulated field 
over the range $2\le l\le 64$. The best fit value of $A$
for WILC9 is found to be $0.090\pm0.029$, 
and for SMICA,  $0.074\pm 0.019$. 
Hence the value matches well with that obtained 
 by hemispherical analysis for the case of SMICA map \cite{Planck2013a}. 
For the case of WMAP \cite{Hoftuft2009}, 
 it is found to be a little larger. 
However the difference 
is not statistically significant. 

As a consistency check we use Eq. \ref{eq:corr_aniso} to determine the
statistic, given our extracted value of $A$. 
 Here we compute the right hand side of this equation
by directly using the $C_l$ values for different multipoles. This can 
be used to obtain an estimate of the statistic over the chosen multipole
range, given a value of the dipole amplitude parameter, $A$. Using the 
value of $A=0.090 \pm 0.029$ for WMAP we obtain the value,  
$S_H = 0.025\pm 0.008$, which is same as that given in Table \ref{tb:dipole1}.
For SMICA the corresponding value is $S_H = 0.020\pm 0.005$,
obtained by using $A= 0.074\pm 0.019$, again
in good agreement with the bias corrected value given in Table \ref{tb:dipole1}.  

Finally we extract the value of the amplitude, $g_0$, of the inhomogeneous
term in the power spectrum which fits data over the entire multipole range,
$2\le l\le 64$.
In this case we set the spectral index $\alpha=0$. Comparison with data,
leads to the value, $g_0 = 0.087\pm0.028$ for WILC9 and
$g_0 = 0.077\pm 0.020$ for SMICA.
This corresponds to the value of $g_0$ for which the theoretical value 
of the statistic matches the data value given in Table \ref{tb:dipole1}.

\begin{figure}[!th]
\centering
\includegraphics[scale=0.70]{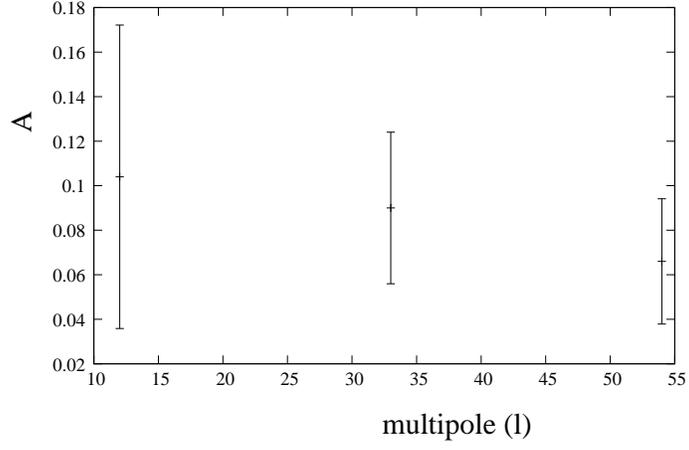}
\caption{The extracted value of the dipole modulation parameter, $A$, 
as a function of the multipole $(l)$, after fixing the direction
parameters, for the three chosen multipole bins, $2-22,\ 23-43,\ 44-64$.
} 
\label{fig:SH_l2}
\end{figure}

\begin{table}[th!]
\begin{tabular}{|c|c|c|c|c|}
\hline
& $S^{data}_H(L) \ (mK^2)$ & $S^{data}_H(L) \ (mK^2)$  & $(l,b)$ & P-value\tabularnewline
&    &   (bias corrected) & & \tabularnewline
\hline
WILC9  & $0.023 \pm 0.007$ & $0.025 \pm 0.008$  & $(227^o,-14^o)$ & $ 0.55\%$\tabularnewline
\hline
SMICA & $0.021 \pm 0.005$  & $0.023 \pm 0.006$ & $(229^o, -16^o)$ & $ 2.6\%$\tabularnewline
\hline
SMICA (in-painted) & $0.027 \pm 0.007$ & $-$ & $(232^o, -12^o)$ & $0.38\% $\tabularnewline
\hline
\end{tabular}
\caption{The extracted value of $S_H^{data}$
in the multipole range $l=2-64$. 
 This value is obtained from data after maximizing $S_H(L)$ with a
search over the direction parameters. The corresponding direction is
given in galactic coordinates. 
The P-value represents the significance of the detected signal, as explained
in text. For the case of SMICA (in-painted) no bias correction is 
required since here we directly use the full sky map.}
\label{tb:dipole1}
\end{table}
We next study the variation of dipole modulation amplitude, $A$, with the 
multipole $l$ using the map, WILC9. We divide data into 3 bins, $l=2-22,23-43$ and $l=44-64$,
for this purpose. In each bin we determine the value of $S_{H}^{data}$
and extract the best fit value of $A(l)$.
For this analysis we found it convenient to fix the direction to be same as that
obtained over the entire multipole range
$2-64$. These direction parameters are also given in Table \ref{tb:dipole1}. Alternatively, 
we may determine the best fit direction parameters, along with the amplitude,
in each bin. We find that, in this case, the direction parameters show a mild 
dependence on the bin.
The results for this case are shown in Table \ref{tb:SH_l4} and Table \ref{tb:SH_l5} 
 for WILC9 and SMICA respectively. 
Since the dependence is relatively small, especially if we ignore the first
bin, we fix the direction to be equal to
the mean direction over the entire range.
We next extract the values of $S_H^{data}$ and $A$ in the three multipole
bins, $l=2-22,23-43,44-64$,
 by repeating the entire procedure, including the bias analysis, 
described in section 
\ref{sec:dataanalysis}.
The resulting bias corrected 
values of $S_{H}^{data}$ for WILC9 in the three bins, 
$l=2-22,23-43$ and $l=44-64$, are found to be $0.0087\pm 0.0057$, $0.0082\pm 
0.0031$ and
$0.0061\pm 0.0026$ respectively. The corresponding values for the
  SMICA map  
 are found to be $0.0084\pm 0.0055$, $0.0063\pm 0.0033$ and 
$0.0066\pm 0.0032$ respectively. 
The extracted values of $A$ are
 plotted in Fig. \ref{fig:SH_l2} for WILC9. 
We find that
$A$ shows a monotonic decrease with $l$.  
The results obtained with the SMICA and the SMICA in-painted map 
agree with those obtained with the WILC9 map within errors.

\begin{table}[!th]
\begin{tabular}{|c|c|c|c|}
\hline
 $multipole$ & $A(l)$ & $S_{H}^{data}(l)$ & $(l,b)$  \tabularnewline
\hline
 $2-22$ &  $0.134\pm 0.068 $ & $0.011\pm0.006$ & $(225^o,-65^o)$  \tabularnewline
\hline
$23-43$ &  $0.101\pm 0.034$ & $0.0092\pm 0.0031$  & $(229^o,5^o)$ \tabularnewline
\hline
$44-64$ &  $0.076\pm 0.028$ & $0.0071\pm 0.0026$ & $(225^o,-25^o)$ \tabularnewline
\hline
\end{tabular}
\caption{The statistic $S_H^{data}$ in the multipole range $2-22,23-43,44-64$ and the corresponding direction parameters for WILC9.
The extracted values of the effective dipole modulation parameter, $A(l)$, for
these three bins are also shown.
}
\label{tb:SH_l4}
\end{table}

\begin{table}[!th]
\begin{tabular}{|c|c|c|c|}
\hline
 $multipole$ & $A(l)$ & $S_{H}^{data}(l)$ & $(l,b)$  \tabularnewline
\hline
 $2-22$ &  $0.122\pm 0.067 $ & $0.0101\pm0.0055$ & $(228^o,-49^o)$  \tabularnewline
\hline
$23-43$ &  $0.074\pm 0.034$ & $0.0073\pm 0.0034$  & $(241^o,3^o)$ \tabularnewline
\hline
$44-64$ &  $0.065\pm 0.027$ & $0.0074\pm 0.0032$ & $(221^o,-37^o)$ \tabularnewline
\hline
\end{tabular}
\caption{The extracted values of $A(l)$, $S_H^{data}$ and  the direction 
parameters for SMICA.
}
\label{tb:SH_l5}
\end{table}

\begin{figure}[!th]
\centering
\includegraphics[scale=0.80,angle=0]{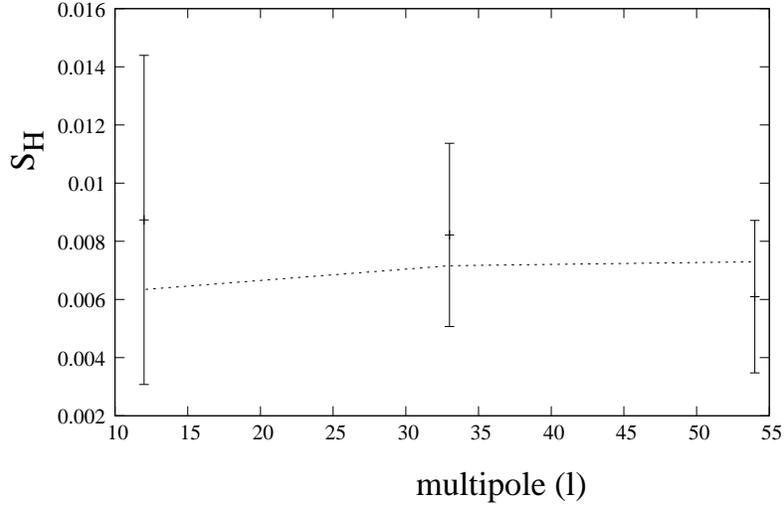}
\caption{The statistic, $S_{H}^{data}$, as a function of the multipole $l$. 
Here the statistic in the three bins is extracted by fixing the direction
parameters to be equal to the mean direction over the entire multipole range.
The dotted line represents the theoretical 
fit corresponding to $\alpha=0, g_0=0.047\pm 0.008$. 
}
\label{fig:SHfit}
\end{figure}

We finally extract the function, $g(k)$, using data in the three
multipole bins, $l=2-22,23-43$ and $l=44-64$. As explained above, the
function, $g(k)$, is parametrized in terms of the overall constant,
$g_0$, and the spectral index $\alpha$. We extract these two parameters
 by minimizing $\chi^2$  
in the three bins, as explained in section
\ref{sec:dataanalysis}.  
We first set $\alpha=0$ and determine $g_0$ which best fits the data in
the three bins. We obtain $g=0.075\pm 0.020$ with $\chi^2=0.50$
 and $g_0=0.070\pm 0.023$ with $\chi^2=0.22$ for WILC9 and SMICA respectively. 
Hence this provides a good fit to data. 
The data clearly favors a zero spectral index for the anisotropic 
part of the power spectrum.   
The resulting fit for WILC9 is shown 
in Fig. \ref{fig:SHfit} as the dotted line.
The value of $\chi^2$ obtained by this fit is relatively small. However
since we are fitting only three data points, such a low value is not
unreasonable especially since the data in the first bin has a relatively large
error. 
Allowing a non-zero value of $\alpha$ we find that the $1\sigma$ limit
on this parameter is $-0.34<\alpha<0.4$ for WILC9 and $-0.24<\alpha<0.39$ for SMICA, that is consistent with zero. 
These limiting values are obtained by allowing $g_0$ as a free parameter. 
The extracted values of $g_0$ and $\alpha$ with SMICA in-painted map are consistent with those quoted above.

\section{Conclusion}
We have extended the results obtained in a recent paper \cite{Pranati2013b}, 
which showed that the dipole modulation model leads to correlations
among spherical harmonic multipoles, $a_{lm}$ and $a^*_{l'm'}$
with $m'=m$ and $l'=l+1$. 
In that paper we defined a statistic, 
$S_H$, which provides a measure of this correlation in a chosen multipole 
range. By making a fit to this statistic in three multipole bins, 
$l=2-22,23-43$ and $l=44-64$, 
we find that the effective dipole modulation parameter $A$ slowly decreases
with the multipole, $l$. 

We propose 
an inhomogeneous power spectrum model and show that it also leads
to a correlation among different multipoles corresponding
to $m'=m$ and $l'=l+1$, as in the 
case of the dipole modulation model. 
The inhomogeneous power spectrum is parameterized
by the function, $g(k)$. We first fit the data by assuming that $g(k)$ follows
a purely scale invariant power spectrum, with $\alpha =0$ in Eq. 
\ref{eq:powergk}.
We determine the value of $g_0$ by making a fit over
the entire multipole range, $2-64$ for WILC9 and also SMICA . 
The best fit value for WILC9 is found to be $g_0 = 0.087\pm0.028$ and 
in case of SMICA the value is $g_0=0.077\pm0.020$.  
We next make a fit over the three multipole bins, as in the case of
the dipole modulation model. 
Again setting $\alpha=0$, we obtain, $g_0=0.075\pm 0.020$, with $\chi^2=0.50$ 
 and $g_0 = 0.070\pm 0.023$, with $\chi^2=0.22$ for WILC9 and SMICA 
respectively.
Hence this
provides a good fit
to data and implies that the value of $\alpha$ is consistent with zero.
Allowing the parameter $\alpha$ to vary,
we find that the one sigma limit of $\alpha$ for WILC9 and SMICA
 is $-0.34<\alpha<0.40$
and $-0.24<\alpha<0.39$ respectively.

\section{Appendix A}
In this appendix we derive Eq. \ref{eq:Allp2}. 
We first change the integration variables 
from $\vec k,\vec k'$ to $\vec k_-,\vec k_+$ in Eq. \ref{eq:Allp1}. The Jacobian of this
transformation is unity. We next integrate Eq. \ref{eq:Allp1} by parts 
over the variable $k_{-z}$. We obtain,  
\begin{equation}
A(l,l') = -(4\pi T_0)^2\int d^3k_- d^3k_+ (-i)^{l-l'+1}
\delta^3(\vec k_-)
g(k_+) {\partial\over \partial k_{-z}}
\left[\Theta_l(k)\Theta_{l'}(k') Y^*_{lm}(\hat k) Y_{l'm'}(\hat k')\right] +
\bar A(l,l') 
\label{eq:Allp3}
\end{equation}
where $\bar A(l,l')$ is the surface term, 
\begin{equation}
\bar A(l,l') = (-i)^{l-l'+1} (4\pi T_0)^2\int  d^3k_+dk_{-x} dk_{-y} B(l,l') 
\label{eq:Allp4}
\end{equation}
and
\begin{equation}
B(l,l') = \int_{-\infty}^\infty  dk_{-z} 
{\partial\over \partial k_{-z}}\left[\delta^3(\vec k_-) g(k_+) 
\Theta_l(k)\Theta_{l'}(k') Y^*_{lm}(\hat k) Y_{l'm'}(\hat k')\right]
\label{eq:Allp5}
\end{equation}
In order to evaluate this we use a standard representation of the 
delta function, given by, 
\begin{equation}
\delta (x) = \lim_{\epsilon\rightarrow 0^+} {1\over 2\sqrt{\pi\epsilon}}
e^{-x^2/(4\epsilon)} 
\end{equation}
We may now express $B(l,l') $ as,
\begin{equation}
B(l,l') = \lim_{\epsilon\rightarrow 0^+} {1\over 2\sqrt{\pi\epsilon}}
 \lim_{K\rightarrow \infty}\int_{-K}^K  dk_{-z} 
{\partial\over \partial k_{-z}}\left[\delta( k_{-x})\delta(k_{-y}) e^{-k_{-z}^2/(4\epsilon)} g(k_+) 
\Theta_l(k)\Theta_{l'}(k') Y^*_{lm}(\hat k) Y_{l'm'}(\hat k')\right]
\label{eq:Allp6}
\end{equation}
As we perform the integral over $k_{-z}$ the term in bracket gets 
evaluated at $k_{-z}^2 = K^2\rightarrow \infty$ at both the upper
and lower limit of the integral. Hence this term is 
proportional to $\exp{(-\infty)}$ which is equal to zero.   
For the remaining term in Eq. \ref{eq:Allp3}, 
we change the integration variables back to $\vec k, \vec k'$. This gives us Eq. 
\ref{eq:Allp2}.

{\bf Acknowledgments:}
Some of the results in this paper have been derived using
the Healpix package \cite{Gorski2005}. We are grateful to James Zibin for a very useful communication and
thank Shamik Ghosh for useful comments.
Finally we acknowledge the use of Planck data available from NASA LAMBDA site
(http://lambda.gsfc.nasa.gov).

\end{document}